\documentclass[aps,prx,twocolumn,amsmath,amssymb]{revtex4} 
\usepackage{graphicx}
\usepackage{graphicx}
\usepackage{Jonasmacros} 
\usepackage{bm}
\usepackage{float}
\usepackage{braket}
\usepackage{mathptmx}
\usepackage{epstopdf} 
\usepackage{txfonts}
\usepackage[colorlinks,citecolor=blue,linkcolor=Fuchsia]{hyperref}
\usepackage[usenames,dvipsnames,svgnames]{xcolor}

\newcommand{\Hbm}{\bm{\mathcal{H}}}
\newcommand{\beq}{\begin{equation}}
\newcommand{\eeq}{\end{equation}}

\newcommand{\ddager}{\bm{d}^{\dagger}}
\newcommand{\dndager}{\bm{d}}
\newcommand{\cdager}{\bm{c}^{\dagger}}
\newcommand{\cndager}{\bm{c}}
\begin{document} 
\date{\today} 
\title{Aharonov Bohm Effect in Voltage Dependent Molecular Spin Dimer Switch} 
\author{Juan David Vasquez Jaramillo, Ph.D}
\email{juan.jaramillo@oist.jp}
\affiliation{Quantum Transport and Electronic Structure Theory Unit,\\
Okinawa Institute for Science and Technology,\\
Okinawa, Japan.}
\author{Erik Sj\"oqvist, Ph.D}
\affiliation{Department of Physics and Astronomy,\\
Uppsala University,\\
Uppsala, Sweden.}
\author{Jonas Fransson, Ph.D}
\email{jonas.fransson@physics.uu.se}
\affiliation{Department of Physics and Astronomy,\\
Uppsala University,\\
Uppsala, Sweden.}
\begin{abstract}
The experimental realization of a coupled spin pair has been reported by Heiko Webber \textit{et.al} and its theoretical description has been previously discussed including the condition that local magnetization of the junction is required for the individual moments to affect the electrons in the molecular ligand through the Kondo interaction. Here in this work, we show that when the couple spin pair is placed in an interferometry set up of the Aharonov-Bohm type additional features related to the switching behavior of the coupled spin pair emerge. This features lead to a phase dependent exchange magnetic field coming from the ferromagnets in proximity with the molecule, a phase dependent commutation of the singlet/triplet ground state around zero bias and it leads to variations in the voltage dependent effective exchange profile between the spin pair. These predictions contribute to the acceptance of the hypothesis that spin polarization can be harvested from quantum coherence in molecular quantum mechanics.
\end{abstract} 

\maketitle
\section{Introduction}
\begin{figure}[h]
	\centering
		\includegraphics[width=0.55\textwidth]{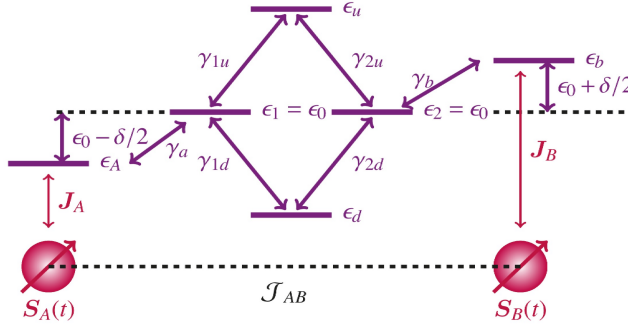}
	\caption{Two Spin RKKY Interfering Machine: Two electronic levels $\epsilon_{A}$ and $\epsilon_{B}$ are coupled to two magnetic units labeled as $\bm{S}_{A}$ and $\bm{S}_{B}$ via Kondo interaction constants $\bm{J}_{A}$ and $\bm{J}_{B}$. Both magnetic units interact through the RKKY interaction where the electron bath is an Aharonov-Bohm like interferometer composed by two electronic beam splitters $\epsilon_{1}$ and $\epsilon_{2}$, and an upper and lower interferometer branch respectively labeled as $\epsilon_{u}$ and $\epsilon_{d}$. A phase $\phi_{1d}$ is printed in the hybridization parameters $\gamma_{1d}$ and $\gamma_{2d}$, which plays the role of an Aharonov-Bohm phase.}
	\label{ABint}
\end{figure}

The control of magnetic interactions at the nanoscale has become essential for writing and reading from a new generation of devices based on single to few spin logic \cite{Mannini2009,Otte2009,Loth2012}, and inducing new effects is of fundamental and practical interest for the scientists working in the field \cite{Bogani2008,Jacobson2015,Jungwirth2016}. As typical reported examples one can point at the control of the single ion uniaxial magnetic anisotropy \cite{Heinrich2015,VasquezJaramillo2018a}, control of the atom by atom magnetic interactions \cite{Khajetoorians2012,Saygun2016a,Jaramillo2017}, single molecule spin dynamics \cite{Fransson2010b,Hammar2016,Hammar2018} and noncollinear order and chiral exchange control on individual atoms \cite{Nino2014,Wu2017}. The typical ways of control include voltage \cite{Fransson2014,Jaramillo2017} and magnetic fields \cite{Khajetoorians2012}, even though it is desired to use less and less magnetic fields, this seems unavoidable at the mean time, and temperature dependent control has been introduced and it has been found useful mainly in magnetic tunnel junctions that operate as thermal diodes \cite{Fransson2014,Jaramillo2017,VasquezJaramillo2017}. On the other hand, as it has been previously shown analytically, a symmetry which is broken by magnetic fields in spin systems is a condition to influence the electronic background through the Kondo interaction, what was overcome by using ferromagnetic leads and the exchange field they produce, hence leaving magnetic fields out of the picture at least externally \cite{Saygun2016a,Jaramillo2017}.
\\
\\
An additional way of control at the nanoscale has been coherence \cite{Aharony2011,Tu2012,Matityahu2013,Tu2014,Liu2016}, up to the extent that it has been claimed that thermodynamic work can be harvest from quantum coherence, in a similar way as when it is claimed that spin polarization can be harvest from quantum coherence \cite{VasquezJaramillo2018}. An alternative way to induce quantum coherence in a molecular system, or what is the same, to induce the ability to exhibit quantum interference is through geometric phases and through the Aharonov-Bohm effect \cite{Fradkin2013}. The deficit of the Aharonov-Bohm phase in this particular set up is, that a magnetic field is required, but nonetheless, its magnitude becomes less important as the electrons in the interferometric path will be insensitive to it, and hence, this becomes an attractive technique to induce quantum coherence.
\\
\begin{figure}[h]
	\centering
		\includegraphics[width=0.5\textwidth]{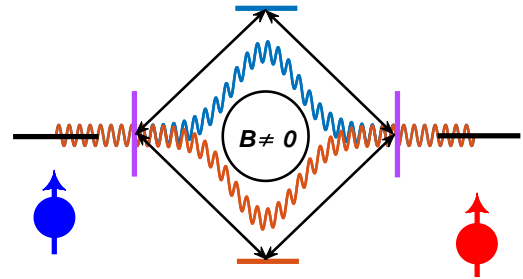}
	\caption{Sketch of the Aharonov-Bohm interferometer. An incoming electron is split in two waves to then interfere with phase $\phi$, in the absence of a magnetic field even though there is a permanent magnetic field in the middle of the ring.}
	\label{sketch_AB}
\end{figure}
Here we propose a molecular Aharonov-Bohm interferometer embedded in a magnetic tunneling junction, where the electronically coupled spins interact only through it as shown in Fig. \ref{sketch_AB}, with its realization as a mesoscopic interferometer shown in Fig. \ref{ABint}. The realization of the mentioned mesoscopic interferometer consists in two energy levels $\epsilon_{a}$ and $\epsilon_{b}$ that play the interchangeable role of electron source and sink which are shifted by $\delta$ energy units with respect to each other, levels $\epsilon_{1}$ and $\epsilon_{2}$ which play the role of beam splitter/recombiner due to their coupling with the two interferometer branches which are resemble by levels $\epsilon_{u}$ and $\epsilon_{d}$. The corresponding couplings between these energy levels giving rise to the mesoscopic interferometer are shown in Fig. \ref{ABint}. This mesoscopic interferometer has as main objective the mediation of the interaction $\mathcal{J}_{ab}$ between spins/magnetic moments $\bm{S}_{a}$ and $\bm{S}_{b}$ through the Kondo interaction with coupling strength $J_{a}$ and $J_{b}$ as shown in Fig. \ref{ABint} resembling the realization by H. Weber \textit{et.al} \cite{Wagner2013}. This molecular structure that play the role of an Aharonov-Bohm interferometer is brought in proximity/contact with two ferromagnetic leads, as shown in Fig. \ref{system}. The basic hypothesis of this work is that, the Aharonov-Bohm phase will play a fundamental role both in the injected spin polarization in the interferometer from the exterior and in the effective interaction between the magnetic units and hence in the spin excitation spectrum and the ground state itself.
\\
The paper is organized as follows: First we detail the system, the model Hamiltonian and the mathematical model to evaluate the nonequilibrium Green's function for the full system in terms of the Aharonov-Bohm Phase and, the effective exchange magnetic field induced from the ferromagnets and the effective isotropic spin-spin interaction (Nonequilibrium RKKY) in terms of the Nonequilibrium Green's function. Second, we establish the most relevant predictions in the part named results, being: 1. The exchange field dependence on the AB phase and coherence degree, 2. The shift in the effective interaction as a function of the degree of coherence and as a function of the AB Phase and 3. the commutation as a function of voltage in the spin excitation spectrum of the coupled spin pair for different AB phases showing the AB Phase dependent ground state shift. Following th section of results, we discuss the more relevant features of the predictions and then conclude. Fruitful appendices on the molecular and nonequilibrium Green's function and the effective theory for the exchange field and the magnetic interactions are presented.
\section{Mathematical Models}
\subsection{Multilevel Molecular Green's Function}
\begin{figure}[h]
	\centering
	\hspace{-3.05cm}
		\includegraphics[width=0.6\textwidth]{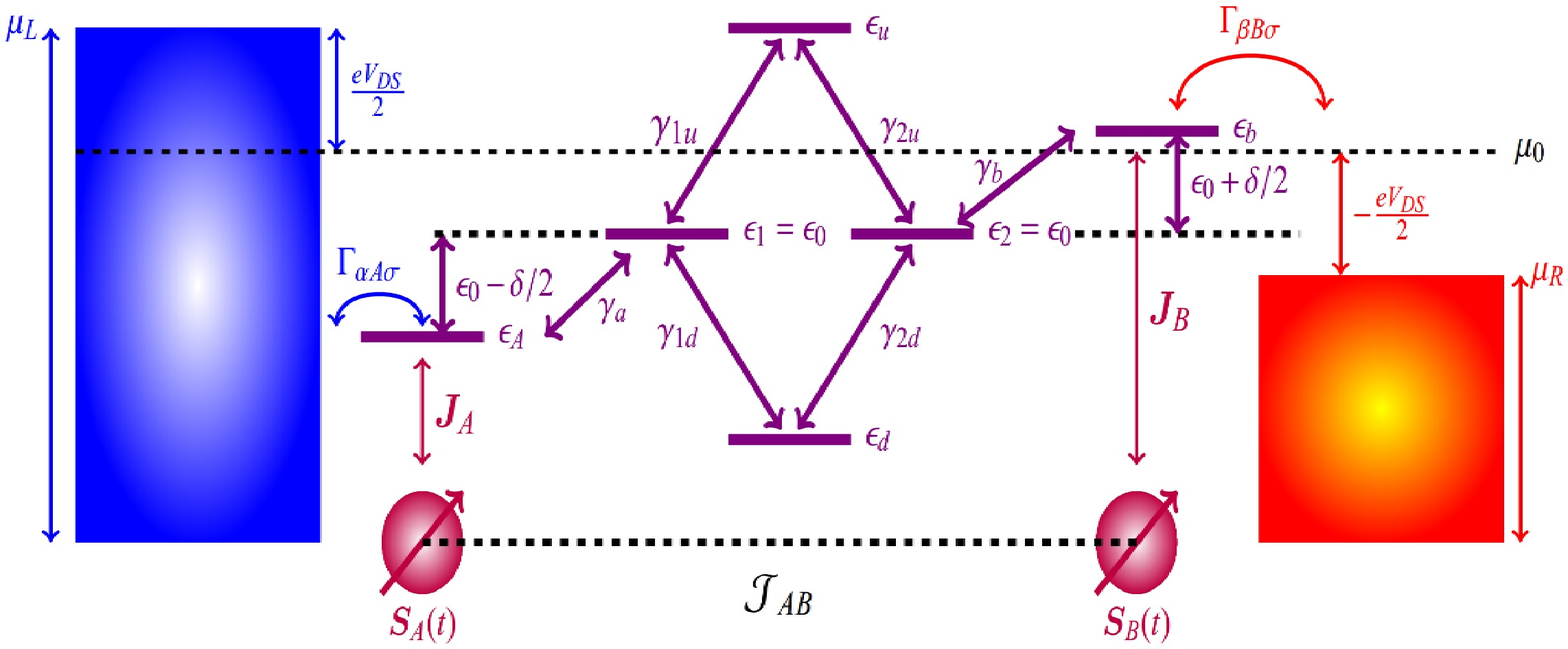}
	\caption{Molecular Magnet Composed by a Dimer of Spins Interacting through a Two Branch AB Electron Interferometer: The couplings to the leads, both left ($\alpha$) and right ($\beta$), are represented by the matrix element $\bm{\Gamma}^{(\alpha,\beta)}_{mn\sigma\sigma}$, }
	\label{system}
\end{figure}

Here we first consider a generalized noninteracting molecule driven out of equilibrium as in reference \cite{VasquezJaramillo2018}, governed by the following model Hamiltonian:
\begin{align}
\Hbm&=\Hbm_{Leads}+\Hbm_{T}+\Hbm_{mol}^{(e)}+\Hbm_{mol}^{(spin)},
\label{app1a}
\\
\Hbm_{Leads}&=\sum_{\bm{k}\sigma}\epsilon_{\bm{k}\sigma}\cdager_{\bm{k}\sigma}\cndager_{\bm{k}\sigma}+\sum_{\bm{q}\sigma}\epsilon_{\bm{q}\sigma}\cdager_{\bm{q}\sigma}\cndager_{\bm{q}\sigma},
\label{app2a}
\\
\Hbm_{T}&=\sum_{m\bm{k}\sigma}V_{m\bm{k}\sigma}\cdager_{\bm{k}\sigma}\dndager_{m\sigma}+V^{*}_{m\bm{k}\sigma}\ddager_{m\sigma}\cndager_{\bm{k}\sigma}
\nonumber
\\
&+\sum_{m\bm{q}\sigma}V_{m\bm{q}\sigma}\cdager_{\bm{q}\sigma}\dndager_{m\sigma}+V^{*}_{m\bm{q}\sigma}\ddager_{m\sigma}\cndager_{\bm{q}\sigma},
\label{app3a}
\\
\Hbm_{mol}^{(e)}&=\sum_{m\sigma}\epsilon_{m\sigma}\ddager_{m\sigma}\dndager_{m\sigma}+\sum_{m_{1}m_{2}\sigma}\gamma_{m_{1}m_{2}\sigma\sigma}\ddager_{m_{1}\sigma}\dndager_{m_{2}\sigma},
\label{app4a}
\\
\Hbm_{mol}^{(spin)}&=\sum_{m\sigma_{1}\sigma_{2}}J_{m}\ddager_{m\sigma_{1}}\bm{\sigma}_{\sigma_{1}\sigma_{2}}\dndager_{m\sigma_{2}}\cdot \bm{S}_{m}(t),
\label{app7a}
\end{align}
where $\dndager_{m\sigma}$ and $\dndager_{m\sigma'}$ creates and annihilates a single electron state in the molecule at the $m-th$ energy level with spin $\sigma$ and spin $\sigma'$ respectively. $\Hbm_{Leads}$ represents the Hamiltonian describing the the metallic leads with band structure specified by $\epsilon_{\bm{k}\sigma}$ and $\epsilon_{\bm{q}\sigma}$ for the left and the right lead respectively, with associated creation and annihilation operators given by $\cndager_{\bm{k,q}\sigma}$ and $\cdager_{\bm{k,q}\sigma}$. The multilevel molecular Hamiltonian here is given by a sum of contributions from the electronic part $\Hbm_{mol}^{(e)}$ and the electron-spin coupling contribution $\Hbm_{mol}^{(e-spin)}$, with $\epsilon_{m\sigma}$ being the level energy, $\gamma_{m_{1}m_{2}\sigma\sigma}$ the level hybridization matrix and $J_{m}$ the Kondo interaction between the spin moment $\bm{S}_{m}(t)$ and the spin of the electrons in the $m-th$ energy level. The nonequilibrium part comes from the coupling between the multilevel molecule and the environment or leads given by the Hamiltonian $\Hbm_{T}$, where the hybridization amplitudes are given by $V_{m\bm{k}\sigma}$ and $V_{m\bm{q}\sigma}$ when hybridizing the $m-th$ energy level of the molecule with the left and right lead respectively. The system driven out of equilibrium is shown in Fig. \ref{system}, where the blue lead represents the colder lead with temperature $T_{L}$, and the red lead represents the hotter lead at temperature $T_{R}$, where in the case of study $T_{L}=T_{R}$. Both chemical potentials, the one in the left lead and the one in the right lead are set to be $\mu_{L}=\mu_{0}+\frac{eV_{DS}}{2}$ and $\mu_{R}=\mu_{0}-\frac{eV_{DS}}{2}$ respectively as labeled in Fig. \ref{system}. Each of the leads is characterized respectively by the Fermi function which is parametrized by the temperature and the chemical potential. Additionally, the couplings to the leads or reservoirs labeled in Fig. \ref{system} as $\bm{\Gamma}_{A\sigma}^{(\alpha)}$ and $\bm{\Gamma}_{B\sigma}^{(\beta)}$, represent the hybridization between levels $\epsilon_{A}$ and $\epsilon_{B}$ respectively with the left and right lead and are related to the self-energy by the expression:
$$
\bm{\Sigma}^{(\chi)}=\bm{\Lambda}^{(\chi)}-\frac{i}{2}\bm{\Gamma}^{(\chi)},
$$ 
and to the hybridization amplitude matrix element between the molecule and the metal through the retarded version of the self-energy:
$$
\Sigma^{R(\chi)}_{mn\sigma\sigma}=\sum_{\bm{k}}\frac{|V_{m\bm{k}\sigma}||V_{n\bm{k}\sigma}|e^{-i(\phi_{m}-\phi_{n})}}{\epsilon-\epsilon_{\bm{k}\sigma}+i\delta}.
$$
Moreover, we derive a generalized formula for a molecular Green's function, and then we specify the tensors that define such object according to our system of study given in Fig. \ref{system}. We depart from the generalized equation of motion given by:
\begin{equation}
i\hbar\frac{\partial \bm{G}_{mn\sigma\sigma'}(t,t')}{\partial t}=\delta_{mn}\delta_{\sigma\sigma'}\delta(t-t')
-\frac{i}{\hbar}\left\langle\bm{T}_{K}\left[\dndager_{m\sigma}(t),\Hbm\right]\ddager_{n\sigma'}(t')\right\rangle,
\label{appD1}
\end{equation}
which is solved in appendix \ref{deriveGreen}. By defining the left and right self-energies correspondingly as follows:
\begin{align}
\bm{\Sigma}^{(L)}_{m\mu \sigma\sigma}(t,t')=\sum_{\bm{k}}V^{*}_{m\bm{k}\sigma}(t)V_{\mu\bm{k}\sigma}(t')\mathcal{G}_{\bm{k}\sigma}(t,t'),
\label{appD13}
\\
\nonumber
\\
\bm{\Sigma}^{(R)}_{m\mu \sigma\sigma}(t,t')=\sum_{\bm{q}}V^{*}_{m\bm{q}\sigma}(t)V_{\mu\bm{q}\sigma}(t')\mathcal{G}_{\bm{q}\sigma}(t,t').
\label{appD14}
\end{align}
\\
where $V_{\mu\bm{k,q}\sigma}(t')$ defines a tunneling process from $m-th$ energy level of the molecule into the left lead for wavevector $\bm{k}$ or into the right lead for wavevector $\bm{q}$ at time $t'$, $V^{*}_{m\bm{k}\sigma}(t)$ defines the tunneling process from the left lead for wavevector $\bm{k}$ or from the right lead for wavevector $\bm{q}$ into the $m-th$ energy level of the molecule, and $\mathcal{G}_{\bm{k,q}\sigma}(t,t')$ is the Green's function for the corresponding lead. Then, the complete solution for the Molecular Nonequilibrium Green's function can be written as follows:
\begin{widetext}
\begin{align}
\bm{G}_{mn\sigma\sigma'}(t,t')=&\int\delta_{mn}\delta_{\sigma\sigma'}\mathcal{G}_{m\sigma}(t,t')
+\sum_{m_{1}}\gamma_{mm_{1}}\int\mathcal{G}_{m\sigma}(t,\tau)\bm{G}_{m_{1}n\sigma\sigma'}(\tau,t')d\tau
+\sum_{\sigma_{1}}J_{m}\int\bm{\sigma}_{\sigma\sigma_{1}}\cdot\left\langle  \bm{S}_{m}(\tau)\right\rangle
\mathcal{G}_{m\sigma}(t,\tau)\bm{G}_{mn\sigma_{1}\sigma'}(\tau,t')d\tau
\nonumber
\\
&
+\sum_{\mu}\int \int\mathcal{G}_{m\sigma}(t,\tau)\bm{\Sigma}^{(L)}_{m\mu \sigma\sigma}(\tau,\tau')\bm{G}_{\mu n\sigma\sigma'}(\tau',t')d\tau'd\tau
+\sum_{\mu}\int \int\mathcal{G}_{m\sigma}(t,\tau)\bm{\Sigma}^{(R)}_{m\mu \sigma\sigma}(\tau,\tau')\bm{G}_{\mu n\sigma\sigma'}(\tau',t')d\tau'd\tau
\label{appD17}
\end{align}
\end{widetext}
where the molecular Green's function for the close system is denoted by $\mathcal{G}_{m\sigma}(t,t')$. Now, we particularize expression \ref{appD17} for the system shown in Fig. \ref{system}. By denoting the Green's function for the open system as $\bm{g}_{0}$, the coupling matrix which will contain the Aharonov Bohm phase as $\left[\gamma\right]$ and the corresponding coupling to the leads as $\Gamma^{(L)}$ and $\Gamma^{(R)}$, we write the contour ordered nonequilibrium stationary Green's function as follows: 
\begin{align}
\bm{G}=\left(\color{violet}\bm{g}_{0}^{-1}\color{black}-\color{violet}\left[\gamma\right]\color{black}+\frac{i}{2}\left(\color{blue}\Gamma^{(L)}\color{black}+\color{red}\Gamma^{(R)}\color{black}\right)\right)^{-1}.
\label{dyson_benzene1}
\end{align}
From the equation of motion given by expression \ref{appD1}, the effect of the Kondo coupling can be determined, as analogous to a Zeeman field, which can be written as:
\begin{align}
\color{violet}\overline{\epsilon}_{a\sigma}=\epsilon_{a\sigma}+\color{purple}J_{a}\left\langle \bm{S}_{a}\right\rangle\sigma_{\sigma\sigma}^{(z)},
\label{dyson_benzene5}
\\
\color{violet}\overline{\epsilon}_{b\sigma}=\epsilon_{b\sigma}+\color{purple}J_{b}\left\langle \bm{S}_{b}\right\rangle\sigma_{\sigma\sigma}^{(z)}.
\label{dyson_benzene6}
\end{align}
from where the bare molecular Green's function can be written as follows:
\begin{align}
\color{violet}
\bm{g}_{0}^{-1}&=\color{violet}\left[
\begin{array}{cccccc}
\hbar\omega-\overline{\epsilon}_{a\sigma}&0&0&0&0&0
\\
0&\hbar\omega-\epsilon_{1\sigma}&0&0&0&0
\\
0&0&\hbar\omega-\epsilon_{d\sigma}&0&0&0
\\
0&0&0&\hbar\omega-\epsilon_{u\sigma}&0&0
\\
0&0&0&0&\hbar\omega-\epsilon_{2\sigma}&0
\\
0&0&0&0&0&\hbar\omega-\overline{\epsilon}_{b\sigma}
\end{array}
\right].
\label{dyson_benzene2}
\end{align}
Next, the coupling matrix can be defined specifically for the Fig. \ref{system}, which is given by:
\begin{align}
\color{violet}\left[\gamma\right]&=\color{violet}\left[
\begin{array}{cccccc}
0&\gamma_{a}&0&0&0&0
\\
\gamma_{a}&0&\gamma_{1d}&\gamma_{1u}&0&0
\\
0&\gamma_{1d}&0&0&\gamma_{2d}&0
\\
0&\gamma_{1u}&0&0&\gamma_{2u}&0
\\
0&0&\gamma_{2d}&\gamma_{2u}&0&\gamma_{b}
\\
0&0&0&0&\gamma_{b}&0
\end{array}
\right]
\label{dyson_benzene3}
\end{align}
for the specific case of this study we assume that $\gamma_{1d}=\gamma_{2d}$, which becomes the control parameter for quantum coherence in the system in a similar was as the length of an interfering path in a photon interferometer. As for the Aharonov-Bohm phase, we assume $\gamma_{1d}\rightarrow\gamma_{1d}e^{i\phi_{0}}$, where $\phi_{0}$ is the Aharonov-Bohm phase.
\\
\\
To evaluate the typical Green's functions we use analytical continuation and hence obtaining the retarded and advanced form of the Green's functions we use the Keldysh equation to determine the form of the nonequilibrium Green's functions given by:
\begin{align}
\bm{G}^{</>}(t,t')=\int\int d\tau_{1}d\tau_{2}\bm{G}^{R}(t,\tau_{1})\bm{\Sigma}^{</>}(\tau_{1},\tau_{2})\bm{G}^{A}(\tau_{2},t'),
\label{k1}
\end{align}
or in the frequency domain:
\begin{align}
\bm{G}^{</>}(\omega)=\bm{G}^{R}(\omega)\bm{\Sigma}^{</>}(\omega)\bm{G}^{A}(\omega),
\label{k2}
\end{align}
where the lesser/greater self energy is given by:
\begin{align}
\bm{\Sigma}^{</>}(\omega)=(\pm i)\sum_{\alpha}f_{\alpha}(\pm\omega)\bm{\Gamma}^{(\alpha)},
\label{k3}
\end{align}
where $\alpha$ indexes the different reservoirs. Note that all of the above are matrices in site and spin space.
\subsection{Effective Theory of Magnetic Interactions}
We consider a nonequilibrium Fermi gas in the absence of spin-orbit coupling with Hamiltonian matrix elements given by $\mathcal{\bm{H}}_{0}$, with diluted magnetic impurities interacting locally with intinerant electrons with a Hamiltonian matrix elements given by $\mathcal{\bm{H}}_{I}$, which is specified by:
\begin{align}
\bm{\mathcal{H}}_{I}[\eta](\bm{r}')=J(\bm{r},\bm{r}')\bm{S}(\bm{r}')\cdot\bm{\sigma}.
\label{k4}
\end{align} 
Given these information, the partition function of the system can be evaluated as a Keldysh path integral given by:
\begin{align}
\mathcal{Z}=\int\int\mathcal{D}[\bar{\psi},\psi]\mathcal{D}\eta e^{iS[\bar{\psi},\psi,\eta]}
\label{k5}
\end{align} 
with Keldysh action given by:
\begin{align}\hspace{-0.75cm}
S[\bar{\psi},\psi,\eta]=\int\int d^{3}\bm{r}d^{3}\bm{r}'\oint_{K}dt\bar{\psi}(\bm{r},t)\left(i\hbar\frac{\partial}{\partial t}-\bm{\mathcal{H}}_{0}-\bm{\mathcal{H}}_{I}[\eta]\right)\psi(\bm{r}',t),
\end{align} 
where $\eta$ is a variable that represents classical spins, phonons or any other classical field, including possible EM waves. Now by integrating the electron fields in expression \ref{k5}, we can define an effective spin action from a new path integral given by:
\begin{align}
\mathcal{Z}=\int\mathcal{D}[\eta]\texttt{Det}\left|(-i)\mathcal{G}^{-1}\left(1-\mathcal{G}\bm{\mathcal{H}}_{I}\right)\right|=
\int\mathcal{D}[\eta]e^{-S_{eff}[\eta]}
\label{k5}
\end{align} 
and from this effective action, a two time pseudo-effective Hamiltonian can be defined in terms of either retarded, advanced or Keldysh susceptibilities \cite{Katsnelson2004,Fransson2010b,Secchi2013,Fransson2017,VasquezJaramillo2018}, but we decide to follow the procedure in \cite{Fransson2017,VasquezJaramillo2018} just considering the retarded susceptibility, hence writting the pseudo Hamiltonian as follows:
\begin{align}
\mathcal{\bm{H}}_{spin}(t,t')&=\sum_{mn}\mathcal{J}_{mn}(t,t')\bm{S}_{m}(t)\cdot\bm{S}_{n}(t')+\bm{T}_{mn}(t,t')\cdot\bm{S}_{m}(t)\times\bm{S}_{n}(t')
\nonumber
\\
&\bm{S}_{m}(t)\cdot \mathbb{I}_{mn}(t,t')\cdot\bm{S}_{n}(t')-\sum_{m}\mu_{B}\bm{S}_{m}(t)\cdot \bm{B}^{eff}_{m}(t),
\end{align}
where $\mathcal{J}_{mn}(t,t')$ is the isotropic exchange interaction which is one of the quantities of interest in this study, $\bm{T}_{mn}(t,t')$ is the effective anisotropic anti-symmetric chiral exchange, $\mathbb{I}_{mn}(t,t')$ is the anisotropic symmetric exchange interaction which contains the uniaxial and the planar anisotropies mediated by electron and $\bm{B}^{eff}_{m}$ is the effective magnetive field induced by the spin assymetry in the Fermi gas, which is the other quantity of interest. The system in Fig. \ref{system} is driven is such a way that the lattice inversion symmetry of the junction makes $\bm{T}_{mn}(t,t')=0$, and by considering each of the spin units a spin half unit, we make $\mathbb{I}_{mn}(t,t')$ unimportant for the specific problem.
\\
\\
Therefore, we can define an effective spin problem in the time invariant regime for Fig. \ref{system} given by:
\begin{align}
\mathcal{\bm{H}}_{spin}=\mathcal{J}_{AB}(t,t')\bm{S}_{A}\cdot\bm{S}_{B}-\mu_{B}\bm{S}_{A}(t)\cdot \bm{B}^{eff}_{A}-\mu_{B}\bm{S}_{B}(t)\cdot \bm{B}^{eff}_{B}.
\label{spin}
\end{align}
Now by defining two new Green's functions $\bm{g}^{</>}$ and $\mathbb{\bm{G}}^{</>}$:
\begin{align}
\bm{g}^{</>}&=\frac{\bm{G}_{\uparrow}^{</>}+\bm{G}_{\downarrow}^{</>}}{2},
\label{def1}
\\
\mathbb{\bm{G}}^{</>}&=\frac{\bm{G}_{\uparrow}^{</>}-\bm{G}_{\downarrow}^{</>}}{2},
\label{def2}
\end{align}
we may write the parameters of the spin Hamiltonian given by expression \ref{spin} in the following way:
\begin{align}
\mathcal{J}_{mn}(\omega)&=\frac{J_{m}J_{n}}{2}\int \int \frac{g_{mn}^{<}(\epsilon)g_{nm}^{>}(\epsilon')-g_{mn}^{>}(\epsilon)g_{nm}^{<}(\epsilon')}{\hbar\omega-\epsilon+\epsilon'}\frac{d\epsilon}{2\pi}\frac{d\epsilon'}{2\pi}
\nonumber
\\
&-\frac{J_{m}J_{n}}{2}\int \int \frac{\mathbb{\bm{G}}_{mn}^{<}(\epsilon)\cdot\mathbb{\bm{G}}_{nm}^{>}(\epsilon')-\mathbb{\bm{G}}_{mn}^{>}(\epsilon)\cdot\mathbb{\bm{G}}_{nm}^{<}(\epsilon')}{\hbar\omega-\epsilon+\epsilon'}\frac{d\epsilon}{2\pi}\frac{d\epsilon'}{2\pi}
\label{param1}
\end{align}
and 
\begin{align}
\bm{B}^{eff}_{m}=J_{m}\Im\texttt{m}\int \frac{d\omega}{2\pi}\mathbb{\bm{G}}_{mm}^{<}(\omega).
\label{param2}
\end{align}
\section{Results and Discussion}
\subsection{Effective Magnetic Field}
\begin{widetext}
\begin{figure}[h]
	\centering
		\includegraphics[width=1.0\textwidth]{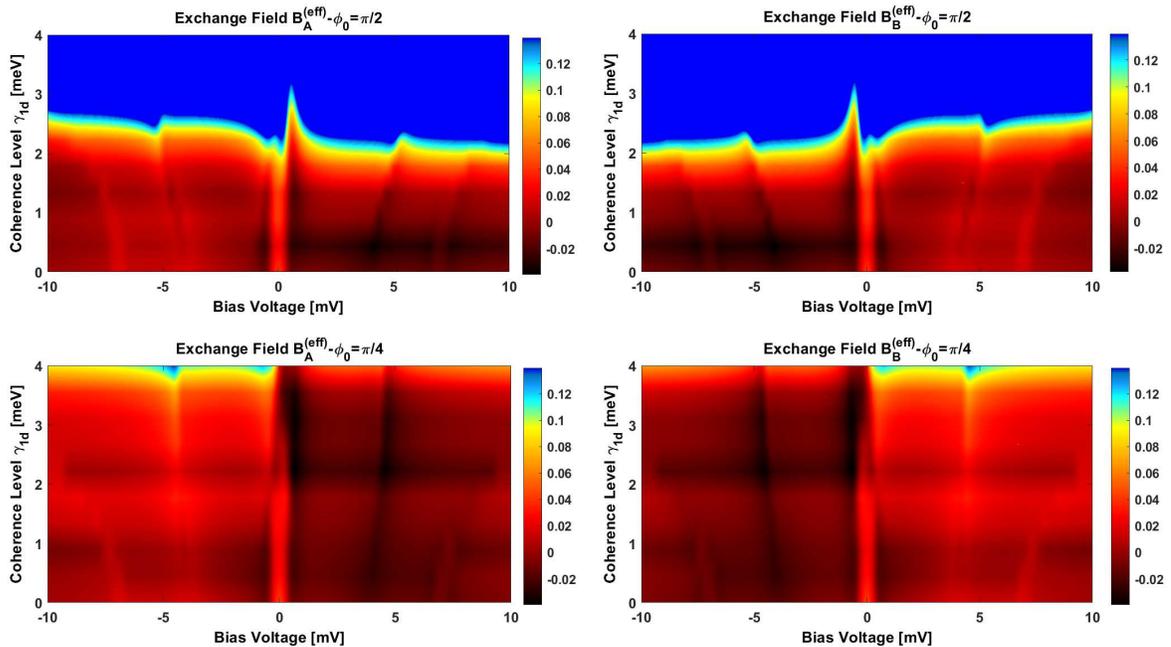}
	\caption{Effective exchange fields: Here we divide the illustration into 4 panels where we show the effective exchange field acting on each spin moment as a function of voltage and coherence control energy for $\phi_{0}=\frac{\pi}{2}$ in the upper two panels, and for $\phi_{0}=\frac{\pi}{4}$ the lower two panels, and for spin labeled as A the left-momst panels and the right most panels for the spin labeled as B.}
	\label{Beff1}
\end{figure}
\end{widetext}

From expression \ref{param2} and expression \ref{def2}, which uses similar notation as the one used in \cite{VasquezJaramillo2018}, one can see that the effective magnetic field or the exchange field, has to do mainly with the question of how efficient can the spin asymmetry from the leads be transfer into the molecule. In this case we find that this spin asymmetry depends heavily on the Aharonov phase and in the degree of coherence as shown in Fig. \ref{Beff1}. 
Each effective magnetic field magnetizes efficiently the junction and hence as shown in previous work \cite{VasquezJaramillo2017,Jaramillo2017}, the nonequilibrium drive will induce a nontrivial behavior on the surrounding electrons and spin moments due to the effective interaction between spins, here given by expression \ref{param1}. For the case shown in Fig. \ref{Beff1}, we show the effective exchange fields acting on both spins due to the surrounding electronic structure for Aharonov-Bohm phases equal to $\pi/2$ and $\pi/4$ respectively due to the fact that modulating the phase in between these values will effectively shift the ground state of the coupled spin pair as shown in Fig. \ref{exchange2}.
\subsection{Shift in the Effective Exchange}
\begin{widetext}
\begin{figure}[h]
	\centering
		\includegraphics[width=1.1\textwidth]{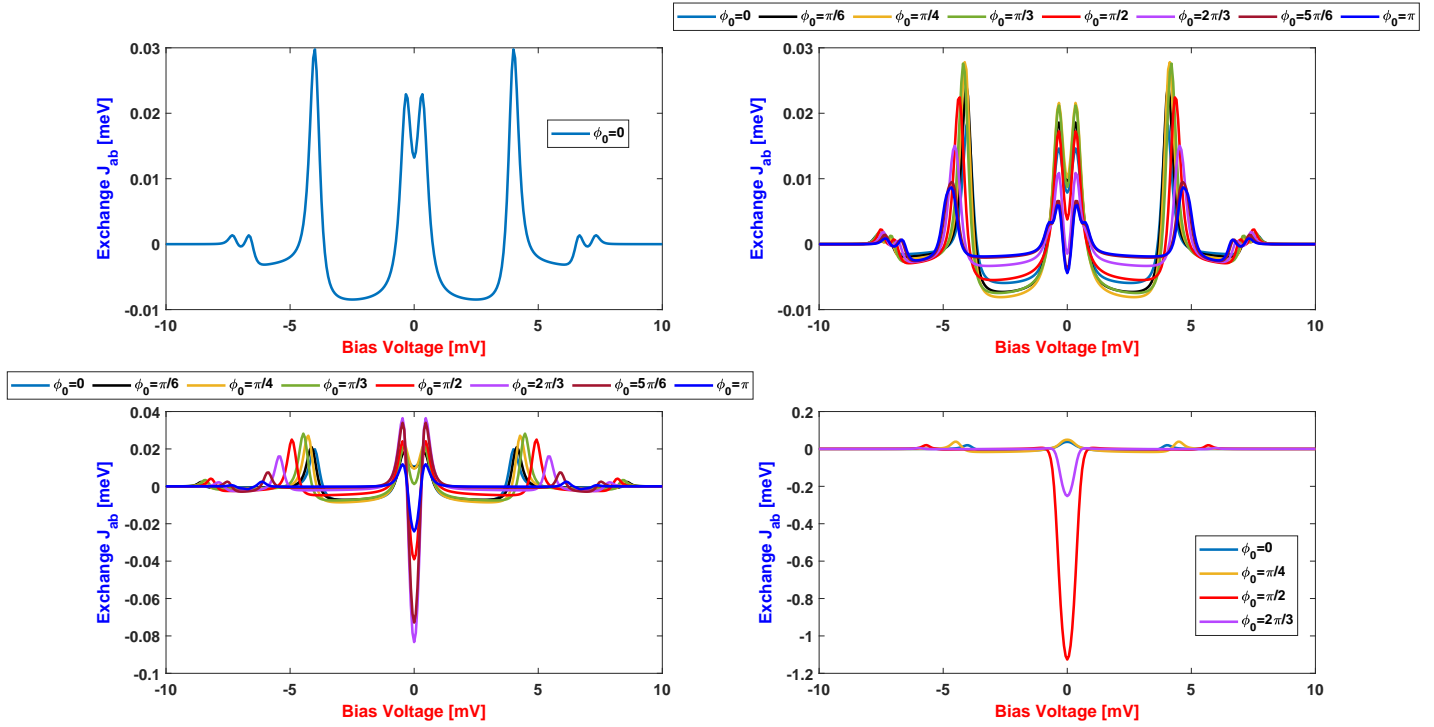}
	\caption{Effective Isotropic Exchange for different values of $\gamma_{1d}$: upper left panel for $\gamma_{1d}=0[meV]$, upper right panel for $\gamma_{1d}=1.8[meV]$, lower left panel for $\gamma_{1d}=3.1[meV]$, and lower right panel for $\gamma_{1d}=4.0[meV]$.}
	\label{exchange1}
\end{figure}
\end{widetext}

\begin{figure}[h]
	\centering
		\includegraphics[width=0.5\textwidth]{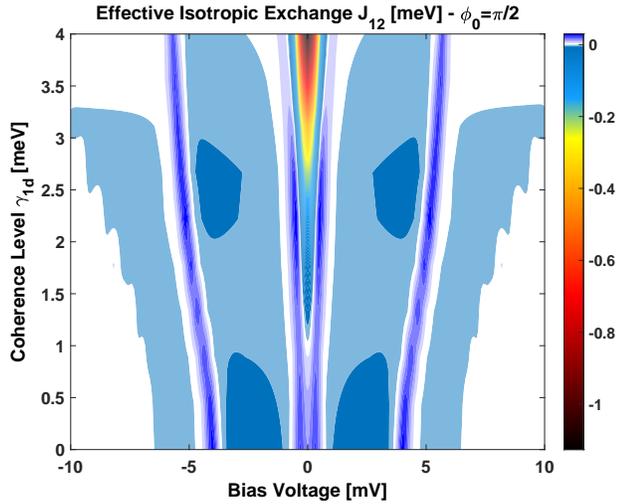}
	\caption{Exchange interaction in the coupled spin pair as a function of voltage and coherence strength.}
	\label{exchange2}
\end{figure}

In previous work we have shown that as a function of voltage, the ground state of a coupled spin pair can be shifted between singlet and triplet, for a spin $\frac{1}{2}$ coupled pair \cite{Jaramillo2017} in experimental agreement with \cite{Wagner2013}. Here, Fig. \ref{exchange1} shows that the Aharonov-Bohm phase is completely capable of commuting the ground state of the coupled spin pair by changing the sign of the effective interaction using phase modulation mechanisms. 
\\
\\
In Fig. \ref{exchange1}, in the upper left panel we can appreciate the normal behavior of the voltage dependent exchange of a system such as the one given by Fig. \ref{system} in the absence of any quantum interference processes. Once quantum interference is allowed by setting $\gamma_{1d}=1.8 [meV]$ times a AB phase factor, a phase dependent shift can be appreciate it in the upper right panel of Fig. \ref{exchange1}. For a value of $\gamma_{1d}=3.1 [meV]$ we can appreciate in the lower left panel of Fig. \ref{exchange1} that for some AB phases the zero bias ground state shift is complete. More importantly, for $\gamma_{1d}=4.0 [meV]$ in the lower right panel of Fig. \ref{exchange1}, we can appreciate a zero bias shift of the ground state of the coupled spin pair by modulating the phase from $\frac{\pi}{2}$ to $\frac{2\pi}{3}$ to $0$ and to $\frac{\pi}{4}$, and for the the initial phases, increasing voltage will induce again a four fold degeneracy in the coupled spin pair contrary to what is observed for phases like $\phi_0=0$ and $\phi_0=\frac{\pi}{4}$, what is shown in the upper left panel. This is one of the key results of the work we are presenting which will be backed up by an analysis of the occupation of each of the ground states for different phases.
Another interesting manifestation of the coherence in the system that goes along with the induced Aharonov Bohm phase is the switching of the exchange interaction with with the change in the coherence strength $\gamma_{1d}$, what has been illustrated in the previous Figure (Fig. \ref{exchange1}), but it can be seen in a greater amount of detail in Fig. \ref{exchange2} for an Aharonov-Bohm phase of $\phi_{0}=\frac{\pi}{2}$. First we consider the important zero bias anomaly presented in Fig. \ref{exchange2}, where around $\gamma_{1d}\approx 1[meV]$ we can observe a shift in sign in the exchange interaction, and hence a shift in the ground state, now clearly because of the coherence strength, which is accompanied by the shift in ground state present when the AB phase is modulated as shown in Fig. \ref{exchange1}. Other important features that can be observed from Fig. \ref{exchange2} are the finite bias features, which exhibit interesting behavior in terms of looking the exchange interaction to zero making the coupled spin pair four fold degenerate for $-5mV\leq V_{DS}\geq 5mV$ for coherence strengths large than $\gamma_{1d}=3.5[meV]$.
\subsection{Spin Occupation and Eigen Energies}
\begin{figure}[h]
	\centering
		\includegraphics[width=0.5\textwidth]{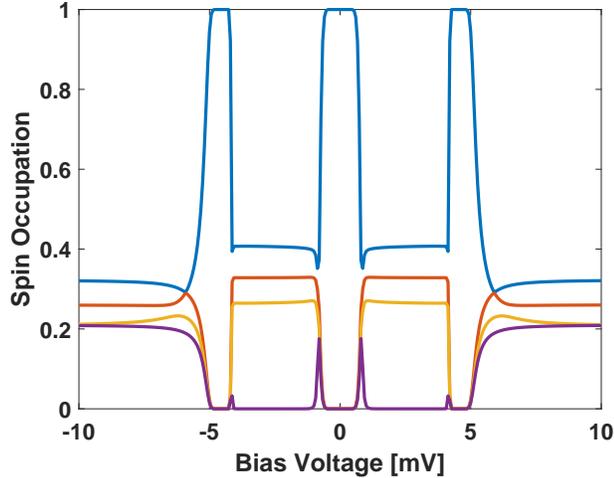}
	\caption{Spin Occupation: Three nearly degenerate states around zero bias and for $-5mV\leq V_{DS}\geq 5mV$, corresponding to a triplet state in the presence of an effective exchange field shown in Fig. \ref{Beff1}. And one state corresponding to a singlet state.}
	\label{occ1}
\end{figure}

\begin{figure}[h]
	\centering
		\includegraphics[width=0.5\textwidth]{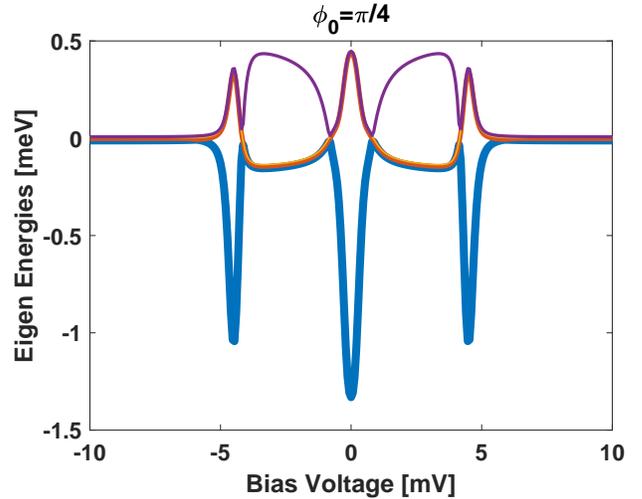}
	\caption{Eigen energy for the case $\phi_{0}=\pi/4$}
	\label{eig2}
\end{figure}

Now we focus on the signatures of the ground state shift in the spin excitation spectrum. A coupled spin pair, in the absence of a magnetic field has a singlet/triplet configuration given by:
\begin{align}
\ket{s}&=\frac{1}{\sqrt{2}}\left(\ket{\uparrow\downarrow}-\ket{\downarrow\uparrow}\right)
\label{spin1}
\\
\ket{t}&=\frac{1}{\sqrt{2}}\left(\ket{\uparrow\downarrow}+\ket{\downarrow\uparrow}\right),\ket{\uparrow\uparrow},\ket{\downarrow\downarrow},
\label{spin2}
\end{align}
where $\ket{s}$ denotes the singlet state with energy $-3\mathcal{J}_{ab}$, and $\ket{t}$ enotes the triplet state which is one with triple degeneracy at energies $\mathcal{J}_{ab}$.
The effect of the magnetic field is noticeable only in the states $\ket{\uparrow\uparrow}$ and $\ket{\downarrow\downarrow}$ which increase and decrease their energy correspondingly, which is signed in to the spin occupation of the coupled spin pair as shown in Fig. \ref{occ1}
Now we focus on the eigen energy plots which will provide support for the claim of ground state shift by modulating the Aharonov Bohm phase in the system shown in Fig. \ref{system}. By looking at Fig. \ref{exchange1} we can see that for an Aharonov Bohm phase of $\pi/4$, the exchange interaction is slightly positive compared to the case when the exchange interaction prominently negative for the case of an Aharonov Bohm phase of $\pi/2$, which leads, according to expressions \ref{spin1} and \ref{spin2} and their corresponding energies to a ground state energy shift under the modulation of the phase. 
\begin{figure}[h]
	\centering
		\includegraphics[width=0.5\textwidth]{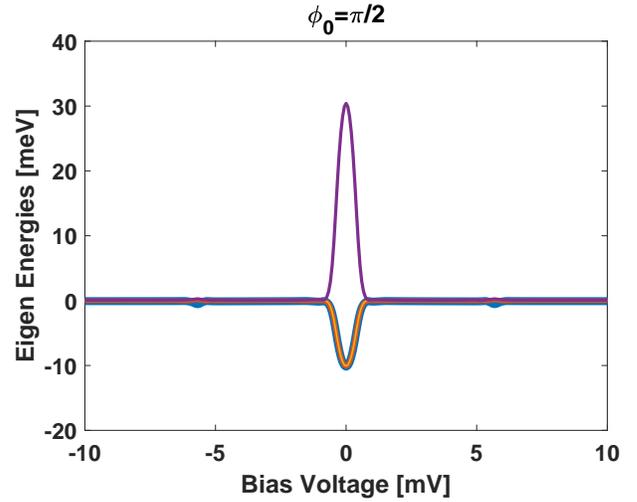}
	\caption{Eigen energy for the case $\phi_{0}=\pi/2$}
	\label{eig1}
\end{figure}
By looking at Fig. \ref{eig2}, we can appreciate a prominent low energy peak of a single state, while a nearly degenerate configuration with three states emerges as a high energy peak, in contrast with Fig. \ref{eig1}, where clearly the triplet state emerges as a lower energy peak, hence, showing a clear signature of a phase modulation based ground state shift.
\section{Conclusions}
Here we presented a coupled spin pair embedded in an electronic Aharonov Bohm like interferometer, which in turn mediates the interaction among the spin moments. Furthermore we drive the interferometer out of equilibrium using a ferromagnetic tunnel junction which produces an exchange field acting on the individual spin moments, being the latter a signature of a spin imbalance or asymmetry in the system injected from the ferromagnetic leads and enhanced through the Aharonov Bohm phase. The degree of coherence of the interferometer is controlled through the hybridization energy $\gamma_{1d}e^{i\phi_{0}}$, being $\phi_{0}$ the Aharonov Bohm phase, and we show that in the presence of an exchange magnetic field, the ground state of the coupled spin pair can be shifted both by modulating the phase and my changing the nature and strength of the quantum coherence in the AB interferometer.
\section{Acknowledgments}
J.D Vasquez-Jaramillo would like to acknowledge financial support from the Colciencias (Colombian Department for Science, Technology and Innovation) 528 Grant for international Doctoral studies and financial support from the Okinawa Institute for Science and Technology (OIST) through the Quantum Transport and Electronic Structure Theory Unit. E. Sj\"oqvist and J. Fransson would like to acknowledge support from Vetenskapr\a det. Authors acknowledge useful discussion with Karlo Penc and Judit Romhanyi.
\bibliographystyle{unsrt}
\bibliography{dimer}
\appendix
\end{document}